\documentclass[prb,showpacs,superscriptaddress, twocolumn]{revtex4}

\usepackage{graphicx,amssymb,amsmath,color,bm}

\begin{document}

\title{Symmetries and the conductance of graphene nanoribbons
with long-range disorder}

\author{J\"urgen Wurm}
\affiliation{Institut f\"ur Theoretische Physik, Universit\"at Regensburg, D-93040 Regensburg, Germany}

\author{Michael Wimmer}
\affiliation{Instituut-Lorentz, Universiteit Leiden, P.O. Box 9506, 2300 RA Leiden, The Netherlands}

\author{Klaus Richter}
\affiliation{Institut f\"ur Theoretische Physik, Universit\"at Regensburg, D-93040 Regensburg, Germany}

\date{\today}

\begin{abstract}
We study the conductance of graphene nanoribbons with
long-range disorder. Due to the absence of intervalley scattering from
the disorder potential, time-reversal symmetry (TRS) can be effectively
broken even without a magnetic field, depending on the type of ribbon
edge. Even though armchair edges generally mix valleys, we show that
metallic armchair nanoribbons possess a hidden pseudovalley structure
and effectively broken TRS. In contrast,
semiconducting armchair nanoribbons inevitably mix valleys and restore
TRS. As a result, in strong disorder metallic
armchair ribbons exhibit a perfectly conducting channel, but
semiconducting armchair ribbons ordinary localization. TRS
is also effectively broken in zigzag nanoribbons in the
absence of valley mixing. However, we show that intervalley scattering
in zigzag ribbons is significantly enhanced and TRS is restored even for
smooth disorder, if the Fermi energy is smaller than the
potential amplitude.
The symmetry properties of disordered nanoribbons are also
reflected in their conductance in the diffusive regime.
In particular, we find suppression of weak localization and an
enhancement of conductance fluctuations in metallic armchair and zigzag ribbons
without valley mixing. In contrast, semiconducting
armchair and zigzag ribbons with valley mixing exhibit weak localization behavior.
\end{abstract}

\pacs{73.63.Nm, 72.80.Vp, 73.20.Fz, 73.23.-b}

\maketitle

\section{Introduction}

The bulk electronic properties of graphene\cite{Castro2009} are modified
significantly in nanoscopic samples, where the influence of the
edges becomes decisive. Edge effects are most prominent in narrow
graphene nanoribbons (GNRs) with the boundary structure determining
whether the electronic spectrum is semiconducting (gapped)
or metallic (gapless).\cite{Tanaka1987, Fujita1996, Nakada1996, Brey2006}
The nanoribbons with the highest symmetry with regard to the
graphene lattice exhibit zigzag and armchair edges, shown in
Fig.~\ref{pic:graphene}. While zigzag GNRs are always metallic,
armchair GNRs are categorized into 'metallic' or 'semiconducting'
depending on their width.\cite{Fujita1996, Nakada1996, Brey2006}
GNRs of an orientation in between armchair and zigzag have been
shown to effectively behave as zigzag GNRs.\cite{Nakada1996,Akhmerov2008}

The first experiments on lithographically defined GNRs failed to find
conclusive evidence for these edge effects,\cite{Han2007,Chen2007} but
since then great effort has been spent on improving the edges of GNRs:
scanning tunneling microscopy has been used to tailor
edges,\cite{Tapaszto2008} GNRs have been chemically derived from
solution phase,\cite{Li2008} they have been obtained by 'unzipping' of
carbon nanotubes,\cite{Kosynkin2009,Jiao2010,Wang2011}
cut out of graphene sheets by anisotropic etching using nickel
clusters\cite{Campos2009} or by sonochemically methods,\cite{Wu2010}
or they were self-assembled inside carbon nanotubes.\cite{Chuvilin2011}
(For an extended overview we refer to a recent review.\cite{Jia2011})

In certain situations, for example when the dynamics
of a system is chaotic or diffusive, its quantum transport
properties are mainly determined by very few symmetries of the
system, in particular the presence or absence of time-reversal
symmetry (TRS).\cite{Beenakker1997} TRS is usually
broken by magnetic fields. However, Berry and Mondragon\cite{Berry1987}
showed that in a (hypothetical) neutrino billiard, TRS
is broken even in the absence of a magnetic field. In fact,
in this case TRS is broken by the edge of the billiard itself.
The Dirac equation used in
Ref.~\onlinecite{Berry1987} to model neutrinos corresponds exactly to the effective low-energy
Hamiltonian of a single valley of graphene, prompting
efforts to realize such an effective TRS
breaking in graphene, for example by different kinds
of disorder\cite{Morpurgo2006,McCann2006} or edges.\cite{Wurm2009}

Zigzag GNRs have also been shown to exhibit this
kind of effective TRS breaking when
only long-range disorder is present, such that the valleys remain uncoupled
and a single-valley Dirac equation description is applicable. In this case,
(effective) TRS is broken by the zigzag edges, placing zigzag GNRs with
long-range disorder into the unitary symmetry class (no TRS).
\cite{Wakabayashi2007} The symmetry class was also shown to influence
the conductance of zigzag GNRs dramatically: For long-range
disorder zigzag GNRs exhibit a perfectly conducting channel (PCC),
i.e.~a minimum of one conductance quantum even in the strongly localized
regime, whereas they show ordinary localization for short-range disorder
that mixes the valleys and restores TRS.\cite{Wakabayashi2007}
In contrast, armchair GNRs were
generally considered to be in the orthogonal symmetry class
corresponding to TRS.\cite{Yamamoto2009,Wakabayashi2009}

In this paper, we investigate carefully the symmetries of graphene
nanoribbons and their effect on the conductance in the
strongly localized and diffusive regime when only long-range
disorder (that does not mix the valleys) is present.
In particular, we show that in contrast to common belief
the symmetry classification of armchair GNRs depends on whether they
are metallic or semiconducting. While semiconducting armchair
GNRs are found to be in the orthogonal symmetry class,
metallic armchair GNRs exhibit a hidden
pseudovalley structure which leads to effective TRS-breaking and
places metallic armchair GNRs into the unitary symmetry class.
This pseudovalley structure also leads to a perfectly conducting
channel in metallic armchair GNRs.

In addition, we show that zigzag GNRs can exhibit an unexpected
and strong source of intervalley scattering, even for long-range,
smooth potentials. When the magnitude of the
disorder potential exceeds the Fermi energy, electron-hole
puddles are formed and valley scattering can be mediated by the edge
state. In this case, TRS is restored in zigzag GNRs and the PCC
vanishes. This puts an additional restriction on the disorder potential
(apart from being long-ranged) in order to realize the
unitary symmetry class in zigzag GNRs.

Weak localization effects in the conductance in the diffusive regime
of extended bulk graphene have been studied
extensively both theoretically and
experimentally.\cite{Suzuura2002,McCann2006,Morozov2006,Tikhonenko2008,
Mucciolo2010} In contrast, we are aware only of a single theoretical work
for (quasi-onedimensional) GNRs that finds weak
localization behavior.\cite{Ortmann2011}
Here we study the quantum transport properties of GNRs in the
diffusive regime systematically, and show that their behavior
is in agreement with their symmetry classifications. In particular,
we find suppression of weak localization in zigzag GNRs without intervalley scattering and metallic armchair
GNRs. In contrast, zigzag GNRs with strong intervalley
scattering and semiconducting armchair GNRs exhibit weak localization behavior. The symmetry
classification also reflects itself in the conductance fluctuations
of the GNRs.

The paper is organized as follows: In Sec.~\ref{sec:model}
we introduce the tight-binding and Dirac Hamiltonian describing
the electronic structure of graphene, and briefly describe our
numerical method. We investigate the valley scattering
properties of a long-range disorder potential in Sec.~\ref{sec:val},
showing that there is an unexpected source of valley scattering
for zigzag GNRs. In Sec.~\ref{sec:syms} we classify the different
types of GNRs according to their symmetry and study their
quantum transport properties. We finally conclude in
Sec.~\ref{sec:concl}.

\section{Graphene Hamiltonian and quantum transport}
\label{sec:model}

\subsection{Hamiltonian}

\begin{figure}
\includegraphics[width=\linewidth]{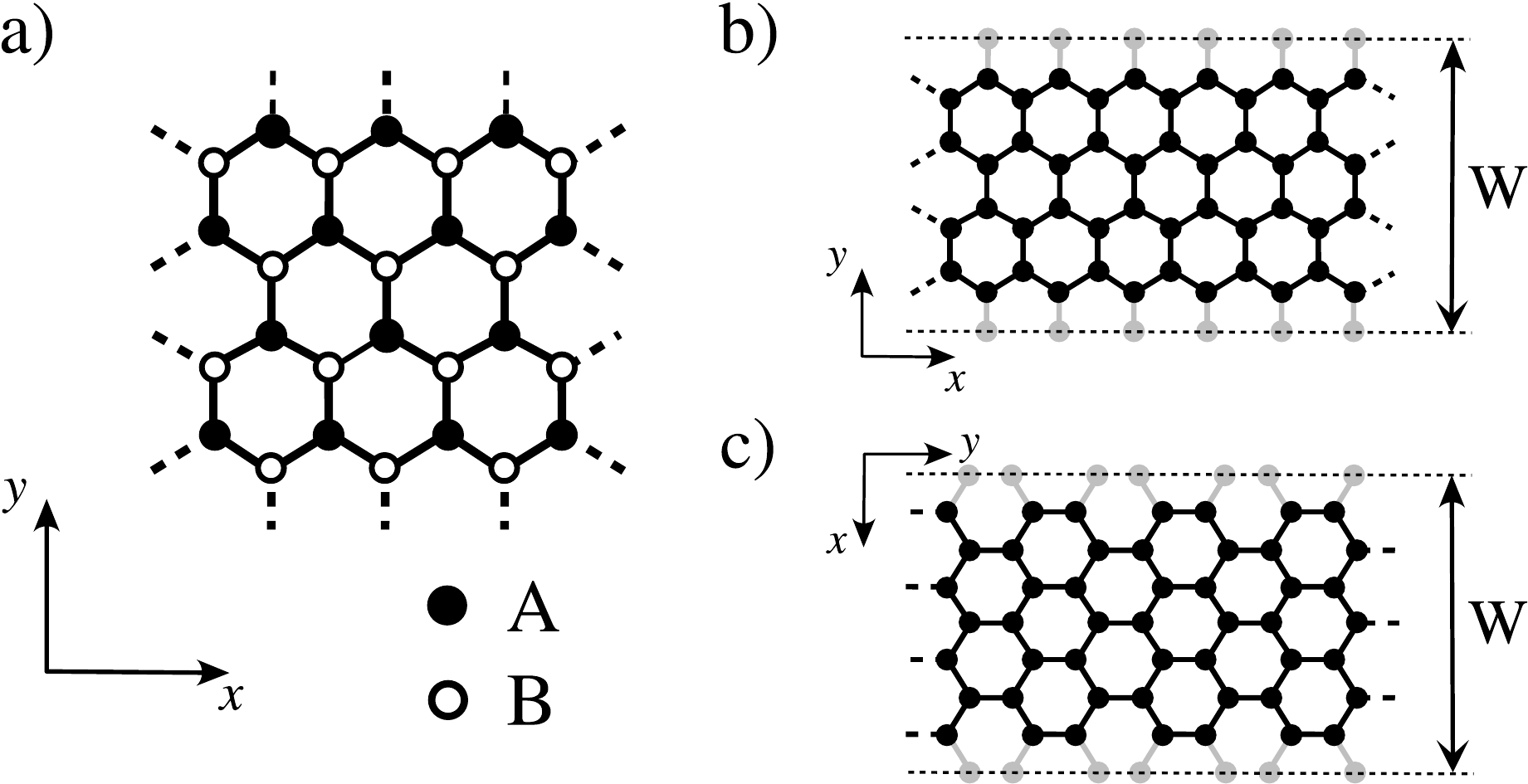}
\caption{(a) The graphene honeycomb lattice. The A and B sublattices
are indicated as solid and open dots, respectively. (b)
zigzag and (c) armchair graphene nanoribbons. The width $W$ of GNRs is
measured between the first rows of missing atoms (shown in grey).
}\label{pic:graphene}
\end{figure}

We describe the electronic structure of graphene using a
tight-binding model for the honeycomb lattice
[shown in Fig.~\ref{pic:graphene}(a)],
\begin{equation}
H = \sum_{i,j} t_{i,j} \left| i\right> \left<j\right| +
\sum_i V(\bm{x}_i)  \left| i\right> \left<i\right|,
\end{equation}
with one orbital $\left| i\right>$ per atom and constant
hopping $t_{ij}=t$ only between nearest neighbors. We allow for an on-site
potential $V(\bm{x})$ that is evaluated at the positions of
the carbon atoms $\bm{x}_i$. A magnetic field is included
through the substitution $t_{ij} \rightarrow
t \times \exp[i\frac{e}{\hbar} \int_{\bm{x}_j}^{\bm{x}_i} d\bm{x}
\bm{A}(\bm{x})]$, where $\bm{A}(\bm{x})$ is the magnetic vector potential.

For a sufficiently smooth potential $V(\bm{x})$ and in the low-energy
limit, excitations with energy $\varepsilon$
obey the Dirac equation
\begin{equation}
H\Psi = \varepsilon\Psi, \label{Dirac_equation}
\end{equation}
where the Hamiltonian
\begin{equation}
H = v_{\rm F}\tau_0\otimes \left(\bm{\sigma}\cdot \bm{p}\right)
+V(\bm{x})\, \tau_0\otimes\sigma_0\label{HDiracdef}
\end{equation}
acts on a four-component spinor wave function
\begin{equation}
\Psi = (\Psi_\text{A},
\Psi_\text{B},-\Psi'_\text{B}, \Psi'_\text{A}).\label{Psidef}
\end{equation}
The Hamiltonian is written in the valley-isotropic form
introduced in Ref.~\onlinecite{Akhmerov2007}:
$\tau_i$ and $\sigma_i$ denote the Pauli matrices in valley and
sublattice space, respectively ($\tau_0$ and $\sigma_0$ are
the respective unit matrices), and $\Psi_j$ and $\Psi'_j$ with
$j \in \{A,B\}$ are the wave function amplitudes on the different sublattices
in the $K$ and $K'$-valley. The Fermi velocity is denoted as $v_\text{F}$
and $\bm{p}=-i\hbar (\partial_x, \partial_y)$ is the two-dimensional
momentum operator, with the orientation of $x$ and $y$-axis as
indicated in Fig.~\ref{pic:graphene}(a). A magnetic field is included through
the minimal coupling $\bm{p} \rightarrow \bm{p}+e\bm{A}(\bm{x})$
with $-e$ the electron charge.

\subsection{Numerical quantum transport in the tight-binding model}

To support our analytical predictions, we perform numerical
computations of the quantum transport properties of
graphene nanoribbons cut out of the graphene lattice
[examples of zigzag and armchair graphene nanoribbons
are shown in Fig.~\ref{pic:graphene}(b) and (c)].

A potential term $V(\bm{x})$ is only introduced in a finite part of the
system (the scattering region), the remaining parts, i.e.
perfect semi-infinite nanoribbons then serve as leads
(with a Fermi energy identical to the scattering region). We  compute the
lattice Green's function of the system using an adaptive recursive Green's
function technique.\cite{Wimmer2009} From the Green's function we then
obtain the scattering matrix using the Fisher-Lee relation for
tight-binding systems.\cite{Sanvito1999} In particular, we compute the
amplitudes $t_{n,m}$ for transmission from mode $m$ to mode $n$ between
two leads, and the amplitudes $r_{n,m}$ for reflection
from mode $m$ to mode $n$ in the same lead. The electrical
conductance $G$ is then obtained using the
Landauer-B\"uttiker formalism\cite{Landauer1957,Buttiker1985},
\begin{equation}
G = G_0 \sum_{n,m} \left| t_{n,m}\right|^2\,,
\end{equation}
where $G_0=2\frac{e^2}{h}$ is the conductance quantum including the spin
degree of freedom.

\section{Inter-Valley scattering in disordered graphene}
\label{sec:val}

\subsection{Impurity potential}

\begin{figure}
\includegraphics[width=\linewidth]{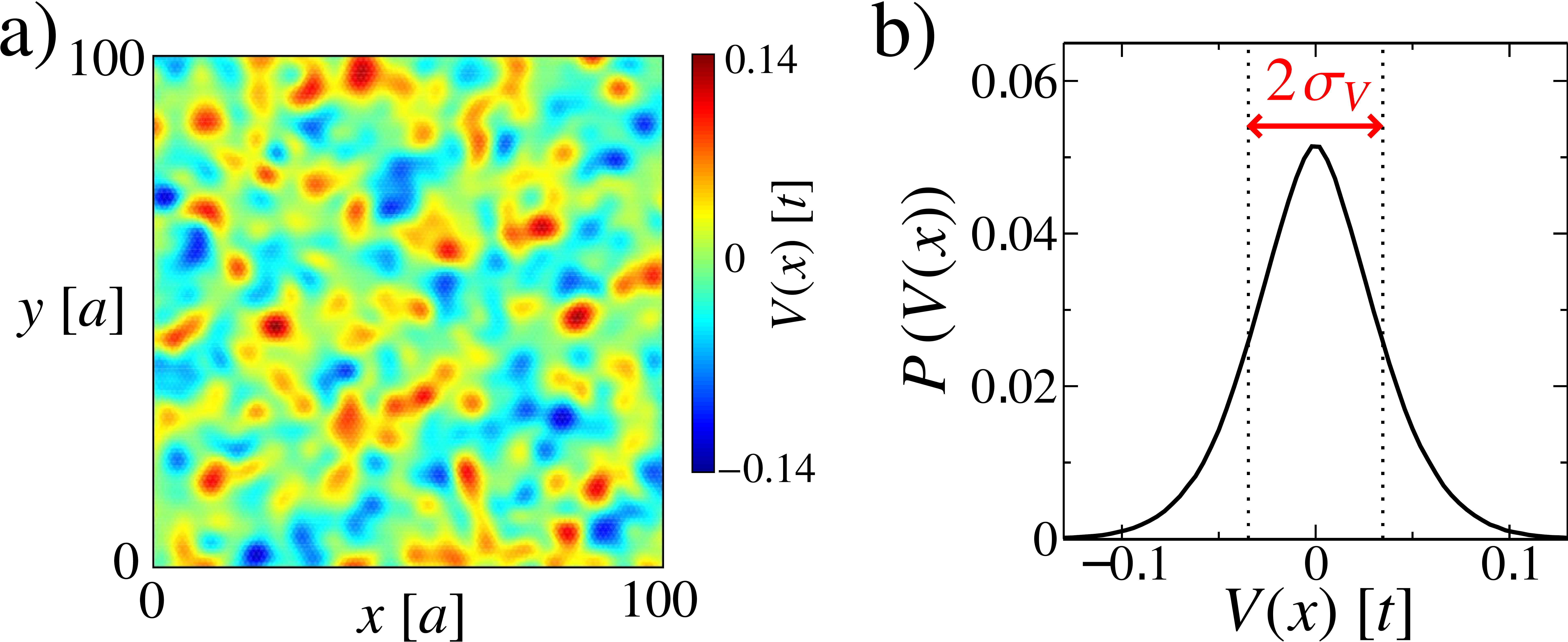}
\caption{(Color online) (a) Example of a realization of the impurity potential
\eqref{eq:dis_pot}. (b) Numerically computed probability distribution of
the value of the potential $V(\bm{x})$ on a given point $\bm{x}$.
For both (a) and (b), $p_\text{imp}=0.05$, $\xi=2a$, and $\delta=0.05\,t$.
}\label{pic:imppot}
\end{figure}

We use a model for an impurity potential that is commonly
used in the study of disordered graphene
(e.g. Refs.~\onlinecite{Rycerz2007,Lewenkopf2008,Krueckl2009,Wurm2010,Mucciolo2010}).
The potential is assumed to consist of a set of independent impurities
with a Gaussian potential profile:
\begin{equation}
\label{eq:dis_pot}
 V(\bm{x}) = \sum_{j=1}^{N_\text{i}} \delta_j \exp\left(-\frac{(\bm{x} - \bm{X}_j)^2}{2\xi^2}\right)\,.
\end{equation}
Here $\bm{X}_j$ is the position of the $j$-th scattering center.
We use $N_\text{i} = p_\text{imp} \,N_\text{a}$ randomly distributed Gaussian
scatterers, where $N_\text{a}$ denotes the total
number of lattice sites in the disordered region, and $p_\text{imp} <1$
is a constant that determines the relative amount of
scatterers. Further, we choose the impurity strength $\delta_j$ randomly
from the interval $[-\delta, \delta]$ and use a constant range
$\xi$ for all impurities.
Figure~\ref{pic:imppot}(a) shows an example of the potential
landscape of this type of impurity potential.

The potential \eqref{eq:dis_pot} can describe both short-range and long-range
impurities by varying the impurity range $\xi$. For $\xi \gtrsim a$, the
resulting potential varies smoothly on the lattice scale, and
the Dirac Hamiltonian \eqref{HDiracdef} is applicable.\cite{Suzuura2002}

The magnitude of the fluctuating impurity potential is best described by its
root mean square value. For $\xi \gg a$ (in practice it is enough to
have $\xi>a$) it is given as
\begin{equation}
\sigma_V = \sqrt{\langle V(\bm{x})^2\rangle} =
\sqrt{\frac{4 \pi\, p_\text{imp}}{3 \sqrt{3}}}\, \delta\, \frac{\xi}{a}\,,
\end{equation}
where $\langle\dots\rangle$ denotes an impurity average.
Figure~\ref{pic:imppot} shows the probability distribution  $P\left(V(\bm{x})\right)$
for finding a particular potential value $V(\bm{x})$ at a point $\bm{x}$. The distribution
is Gaussian-like, with a width given by $\sigma_V$.

In the limit of $\lambda_\text{F} \gg \xi \gtrsim a$ (with
$\lambda_\text{F}$ the Fermi wave length), the transport mean free
path is given in Born approximation as\cite{Suzuura2002, Rycerz2007}
\begin{subequations}\label{eq:ltr}
\begin{equation}
l_\text{tr}= \frac{4}{k_\text{F} K_0}=\frac{2 \sqrt{3} t}{E K_0} a
\end{equation}
with the dimensionless correlator
\begin{equation}
K_0=\frac{4 \pi}{(\hbar v_\text{F})^2} \sigma_V^2 \xi^2\,.
\end{equation}
\end{subequations}

\subsection{Bulk graphene versus zigzag nanoribbons}
\label{sec:valleyscatt}

The parameter $\xi$ of the impurity potential \eqref{eq:dis_pot}
determines the smoothness of the potential. It is generally accepted
that for $\xi \gtrsim  a$ there is only little intervalley scattering.
However, the evidence for this was always only
indirect,\cite{Wakabayashi2007,Rycerz2007,Lewenkopf2008,Mucciolo2010} and no
\emph{quantitative} discussion of intervalley scattering does
exist for this type of potential. Recently,
it has only been attempted to quantify the intervalley scattering
for short-range lattice defects.\cite{Libisch2011}
Since the knowledge of the degree of intervalley scattering will
be important in the following section, we first investigate the
intervalley scattering for the impurity potential \eqref{eq:dis_pot}.
Our findings show that caution must be exerted, as the presence
of zigzag edges may lead to enhanced intervalley scattering even if the impurity
potential is very smooth.

We can numerically measure the intervalley scattering if we consider
a wire along the x-direction, using either periodic boundary
conditions in $y$-direction (making the system equivalent to
an armchair carbon nanotube) or zigzag boundaries. In both cases
the valleys $K$ and $K'$ project onto different values of the
longitudinal momentum $k_x$. The scattering states in the
leads have a definite Bloch momentum $k_x$, and a mode $m$ can thus
be uniquely assigned to a valley (this is not possible for armchair
ribbons, where the two valleys project on the same momentum).
It is then possible to decompose the numerically computed
transmission and reflection probabilities into an
intravalley and intervalley part. The total
probability of reflection into the other valley is given as
\begin{equation}
R_\text{inter}=\sum_{\substack{n \in K, m \in K'\\
    m \in K, n \in K'}} \left|r_{n,m}\right|^2\,,
\label{R-inter}
\end{equation}
while the total reflection probability is given as
$R=\sum_{n,m}\left|r_{n,m}\right|^2$. The intervalley transmission probability
can be defined analogously.

\begin{figure}
\includegraphics[width=\linewidth]{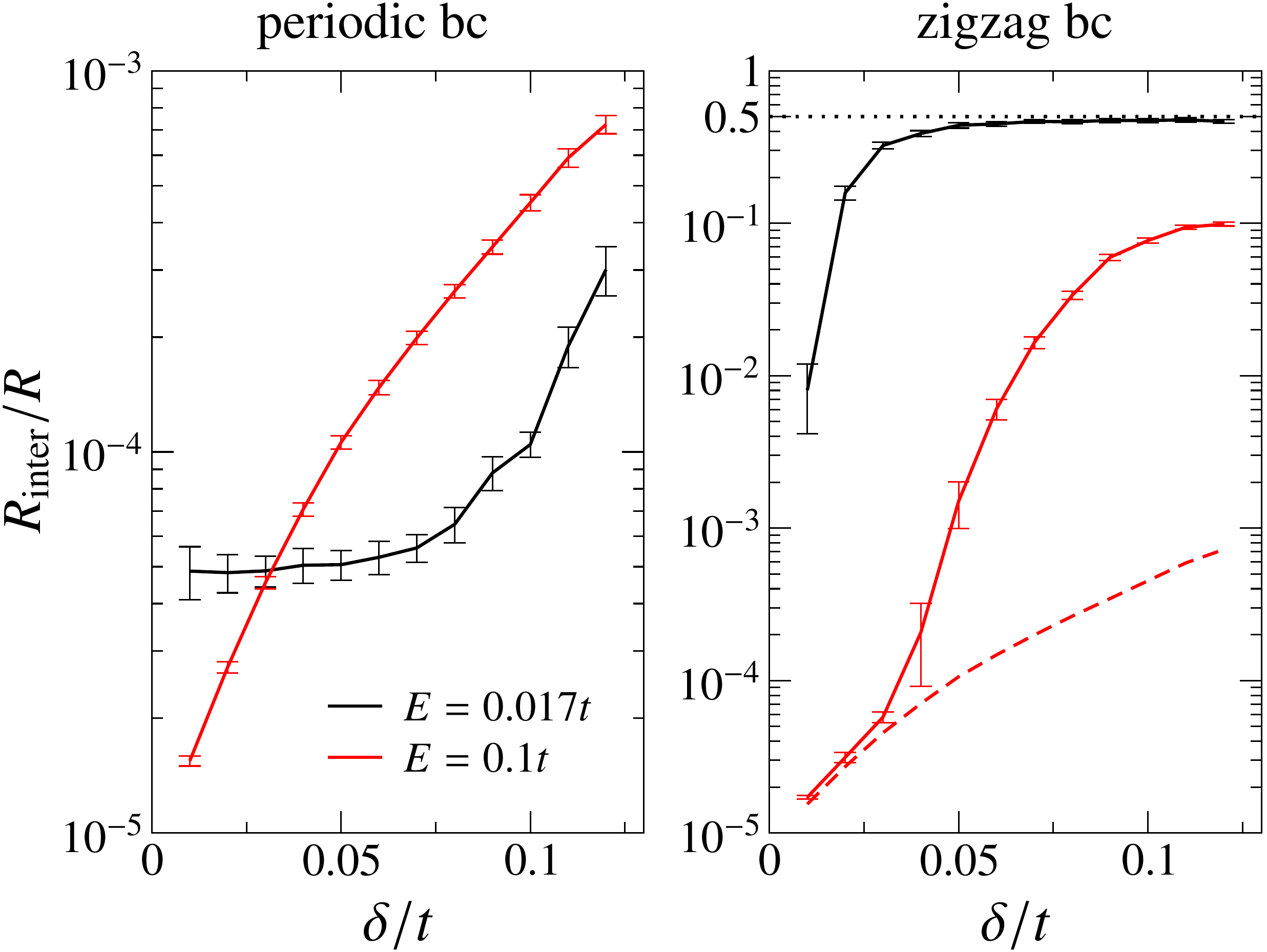}
\caption{(Color online) Probability for intervalley scattering in
reflection [Eq.~(\ref{R-inter})] as a function of impurity strength $\delta$ for
a ribbon with periodic boundary conditions (left panel)
and zigzag boundaries (right panel). The ribbons have width
$W\approx 500a$ and length $L\approx2500a$, and data is shown for
$E=0.017t$ (black solid lines) and $E=0.1t$ (red solid lines).
For comparison, the right panel (with zigzag
boundaries) also contains the data for $E=0.1t$ with periodic
boundary conditions as a red dashed line. The data was obtained
by averaging over 100 different impurity configurations with
potential parameters $p_\text{imp}=0.05$ and $\xi=2a$.}\label{pic:valleys}
\end{figure}

In Fig.~\ref{pic:valleys} we show the fraction of intervalley scattering
in the reflection probability, $R_\text{inter}/R$, as a function
of the impurity strength $\delta$. The rationale for measuring the
intervalley scattering in the reflection is that transmission may
contain a sizeable part of ballistic processes without any scattering;
in contrast, reflection only occurs after at least one scattering
event.
For periodic boundary conditions and $\xi=2a$ we indeed find
only very little intervalley scattering
(left panel of Fig.~\ref{pic:valleys}).
The intervalley
scattering rises with increasing impurity strength $\delta$. This is
to be expected, as for fixed $\xi$ the potential becomes steeper
as $\delta$ is increased, and hence intervalley scattering becomes
more likely. Nevertheless, for the given parameters, the fraction
of intervalley scattering remained below $10^{-3}$.

However, we obtain a very different picture for zigzag graphene nanoribbons
(right panel of Fig.~\ref{pic:valleys}).
Although we use the same impurity potential as in the
case of periodic boundary conditions, we find an intervalley scattering
that is \emph{3 orders of magnitude larger}. The only
obvious difference is the presence of the zigzag edge. Indeed,
it was shown that even a smooth pn-junction in a zigzag nanoribbon
strongly scatters valleys, as both valleys are connected by the
edge state.\cite{Akhmerov2008a} We believe that our numerical findings
can be explained fully by the fact that the impurity potential locally leads
to many smooth pn-junctions (in the bulk, but also at the zigzag edge),
when its magnitude becomes larger than the Fermi energy $E$: For $E=0.1t$
and small $\delta$, the intervalley scattering for the zigzag case
follows the result obtained with periodic boundary conditions (shown
for comparison as a dashed line in the right panel of Fig.~\ref{pic:valleys}).
For larger $\delta$, when the maxima of the potential become
comparable or greater to $E=0.1t$ (which is already the case around
$\delta=0.05\,t$ as seen from Fig.~\ref{pic:imppot}), the
intervalley scattering rate raises rapidly and reaches up to $10\%$. For
smaller Fermi energy $E=0.017t$, this regime is reached already for much
smaller $\delta$, and we find perfect valley mixing close
to $50\%$.

Hence, although the impurity potential \emph{itself} indeed does not scatter
valleys significantly if $\xi \gtrsim a$, caution must be exerted
if zigzag edges are present. Intervalley scattering can be
very large if the impurity potential locally crosses the Dirac point
and hence locally forms pn-junctions. We believe that this unexpected valley
scattering may also explain many not understood
numerical findings of the past.

\section{Symmetry and the conductance of graphene nanoribbons}
\label{sec:syms}

\subsection{Symmetries and quantum transport in
disordered wires}

The quantum transport properties of disordered quantum wires are universal and
determined by their symmetries only.\cite{Beenakker1997} In particular,
the symmetry class is determined by the presence or
absence of time-reversal symmetry (TRS) $\mathcal{T}$, which is an antiunitary
symmetry.
A system may belong to one of the three Wigner symmetry classes:
unitary if TRS is broken, and orthogonal or symplectic if
TRS is present with $\mathcal{T}^2=+1$ or
$\mathcal{T}^2=-1$, respectively.
A system with TRS obeys $\mathcal{T} H \mathcal{T}^{-1}=H$.

Note that a Hamiltonian may possess a TRS that is,
however, irrelevant: This is the case if the Hamiltonian decomposes into
independent blocks and the symmetry connects only between them. In this
case the TRS has no influence on the quantum transport properties
(except guaranteeing a degeneracy between the blocks), and the
symmetry class is determined by intrablock symmetries only. Below,
we identify the appropriate symmetries for the case of graphene
nanoribbons.

\subsection{Symmetries of graphene nanoribbons within the Dirac approximation}
\label{sec:sym}

\subsubsection{Bulk symmetries of graphene and boundary conditions}

The bulk Dirac Hamiltonian \eqref{HDiracdef} commutes with four
antiunitary symmetries\cite{Suzuura2002,McCann2006,Ostrovsky2007,Wurm2009}
\begin{equation}\label{eq:antiunitsym}
\mathcal{T}_i=\tau_i \otimes \sigma_y \mathcal{C}\quad
\text{for $i \in {0,x,y,z}$,}
\end{equation}
where $\mathcal{C}$ denotes complex conjugation. Each of these
antiunitary symmetries can play the role of an (effective) TRS.

It is easy to see that $\mathcal{T}_y^2=1$, whereas $\mathcal{T}_i^2=-1$
for $i \in \{0, x, z\}$. $\mathcal{T}_y$ represents the (true)
TRS that connects the two valleys. $\mathcal{T}_x$ is
the valley symmetry that guarantees the Kramer's degeneracy of both valleys
(since $\mathcal{T}_x^2=-1$).
It should be noted that $\mathcal{T}_0$
and $\mathcal{T}_z$ only differ by a phase in the two valleys and are thus
equivalent. In fact, one can write down a whole family of
equivalent effective intravalley TRSs\cite{moregeneralt0z}
\begin{equation}
\mathcal{T}_{0z}(\vartheta)=\left( \cos\vartheta\, \tau_0 +
i \sin\vartheta\, \tau_z \right) \otimes \sigma_y \mathcal{C} \,.
\end{equation}
From this family, a single antiunitary symmetry will survive in the case
of metallic armchair nanoribbons, as we will show below. The presence
of a magnetic field breaks all four symmetries $\mathcal{T}_i$.

The other important symmetries of graphene, chiral symmetry $(\tau_z\otimes\sigma_z) H
(\tau_z\otimes\sigma_z)=-H$, and particle-hole symmetry
$(\tau_0\otimes\sigma_x) H^* (\tau_0\otimes\sigma_x) = - H$, are broken
by the potential term $V(\bm{x})$. Hence, we do not expect to see universality
classes beyond the three Wigner classes.\cite{Evers2008}

It is well-known that bulk graphene with long-range scatterers belongs to the
symplectic symmetry class.\cite{Suzuura2002,McCann2006} In this case
valleys are not mixed and the true TRS $\mathcal{T}_y$
as well as the valley symmetry $\mathcal{T}_x$ are irrelevant, and the effective
intravalley TRS $\mathcal{T}_{0,z}$ determines the symmetry
class.

In a graphene nanoribbon, the antiunitary symmetries $\mathcal{T}_i$ must also
be compatible with the boundary conditions. Boundary conditions in the
Dirac equation can be written generally in the
form\cite{McCann2004,Akhmerov2008}
\begin{equation}
\Psi(\bm{x})=\mathcal{M}_b \Psi(\bm{x})\quad\text{for $\bm{x}$ on boundary $b$,}
\end{equation}
where $\mathcal{M}_b$ is a Hermitian $4\times 4$-matrix. A graphene nanoribbon is
then symmetric with respect to $\mathcal{T}_i$, if
\begin{equation}
[\mathcal{M}_b, \mathcal{T}_i] = 0\,,
\end{equation}
where $[A,B]=AB-BA$ denotes the commutator.
Below we now specialize to the cases of zigzag and armchair nanoribbons.

\subsubsection{Zigzag nanoribbons}

The boundary condition for a zigzag GNR reads\cite{Brey2006,Akhmerov2008}
\begin{equation}
\mathcal{M}_b=\pm \tau_z \otimes \sigma_z\quad\text{for $b=1,2$.}
\end{equation}
This boundary condition does not mix valleys which thus remain a good
quantum number. The system is symmetric with respect to $\mathcal{T}_y$ and
$\mathcal{T}_x$, but the boundary condition breaks $\mathcal{T}_{0,z}$.
However, since valleys are not mixed for long-range impurities,
the valley-offdiagonal symmetries $\mathcal{T}_{x,y}$ are not relevant. Since
all intravalley TRS are broken by the boundary conditions, a zigzag graphene
nanoribbon resides in the unitary symmetry class, as shown in
Ref.~\onlinecite{Wakabayashi2007}.

These considerations hold as long as there is no intervalley
scattering due to local pn-junctions at the zigzag edge. If there
is, the valleys are strongly mixed, and the only remaining symmetry
is the TRS of the tight-binding lattice ($\mathcal{T}_\text{tb}=\mathcal{C}$,
$\mathcal{T}_\text{tb}^2=1$).
The Dirac equation cannot capture scattering between the
valleys via the edge state since a continuous equation cannot
represent the finite size of the zigzag GNR Brillouin zone
that is at the heart of the scattering mechanism.\cite{Akhmerov2008a}
The zigzag GNR is then in the orthogonal symmetry class.

\subsubsection{Metallic armchair nanoribbons}

The boundary condition for an armchair GNR reads\cite{Brey2006,Akhmerov2008}
\begin{equation}
\label{eq:ac-bc}
\mathcal{M}_b=\bm{\nu}_b\cdot\bm{\tau} \otimes \bm{t}\cdot \bm{\sigma}
\quad ; \quad \bm{\nu}_b=(\sin\vartheta_b, \cos\vartheta_b,0)
\end{equation}
with  $b=1,2$,
and $\bm{t}$ points in the direction of the GNR. This boundary condition
strongly mixes valleys, and the relative valley-angle $\vartheta_2-\vartheta_1$
between the two boundaries of the armchair GNR depends on the width of
the ribbon. For example, for an armchair GNR in $y$-direction
as shown in Fig.~\ref{pic:graphene}(c), $\vartheta_b=-2 K x_b$ where
$K=4\pi/3 a$.\cite{Wurm2011}

A metallic armchair nanoribbon has $\vartheta_1=\vartheta_2=\vartheta$, i.e.
\begin{equation}
\mathcal{M}_1=\mathcal{M}_2=\mathcal{M}\,.\label{eq:ac_bc}
\end{equation}
In this case, $\bm{\nu}\cdot\bm{\tau}$ commutes with both
the Hamiltonian (including disorder) and the boundary condition, and we may
choose the solutions of the Dirac equation as eigenstates of
$\bm{\nu}\cdot\bm{\tau}$. The solutions can thus be grouped into two
new \emph{pseudovalleys}, $K_\mathcal{R}$ and $K'_\mathcal{R}$,
that \emph{remain uncoupled for long-range disorder}. The pseudovalley
description is obtained from the usual valleys by means of the
rotation
\begin{equation}
\mathcal{R}=e^{-i\pi \tau_x/4}e^{-i\vartheta \tau_z/2} \,.
\label{eq-R}
\end{equation}

The metallic armchair boundary condition \eqref{eq:ac_bc} is symmetric
with respect to $\mathcal{T}_y$ and $\mathcal{T}_{0z}(\vartheta)$ with
$\vartheta$ equal to the valley-angle of the boundary condition, whereas
$\mathcal{T}_x$ is broken. In the pseudovalley space they take the form
\begin{equation}
\begin{split}
&\mathcal{T}_y^\mathcal{R} = \tau_y \otimes \sigma_y \mathcal{C}\,,\\
&\mathcal{T}_{0z}^\mathcal{R}(\vartheta) = -i \tau_x \otimes \sigma_y\mathcal{C}\,,
\end{split}
\end{equation}
where $\mathcal{T}^\mathcal{R}=\mathcal{R T R}^\dagger$. Hence, both
symmetries are completely offdiagonal in valley space and are not relevant for
determining the symmetry class in the case of long-range potential. In the
absence of any intra-pseudovalley TRS, metallic armchair GNRs \emph{also
belong to the unitary symmetry class}. Note that Ref.~\onlinecite{Luo2009}
already discussed effective TRS breaking in the context of
the lowest mode of metallic armchair GNRs in a ring geometry. Our analysis
shows the more general result that, in the low-energy limit, metallic armchair
GNRs belong to the unitary symmetry class without a special geometry and
regardless of the number of modes. We will confirm this using numerical
simulations below.

\begin{figure}
\includegraphics[width=\linewidth]{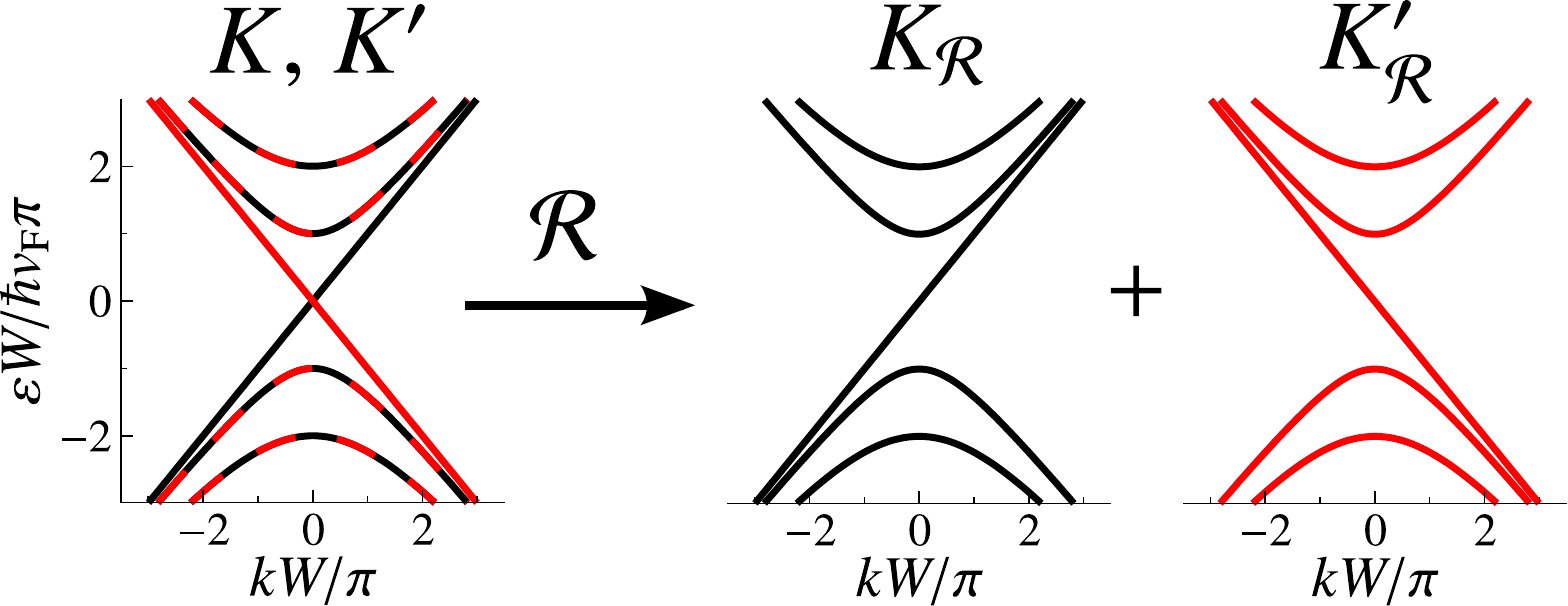}
\caption{(Color online) Pseudovalley resolved band structure of a metallic armchair
nanoribbon after rotation $\mathcal{R}$, see Eq.~(\ref{eq-R}).}\label{pic:acmvalleys}
\end{figure}

A metallic armchair GNR features a band structure $\varepsilon(k)$ (with $k$ the
Bloch wave vector) with
two non-degenerate gapless linearly dispersing bands and pairs of two-fold
degenerate hyperbolic bands (shown in Fig.~\ref{pic:acmvalleys}).\cite{Brey2006}
We can use the conserved antiunitary symmetries to unravel the pseudovalley
structure: Both $\mathcal{T}_y^\mathcal{R}$ and
$\mathcal{T}_{0z}^\mathcal{R}(\vartheta)$ lead to\cite{notepseudovalley}
\begin{equation}
\varepsilon_{K_\mathcal{R}}(k)=\varepsilon_{K'_\mathcal{R}}(-k)\,.
\end{equation}
Hence, the two counter-propagating gapless linear modes are (Kramer's) partners
residing in \emph{different} pseudovalleys. In addition,
every pseudovalley contains a set of non-degenerate hyperbolic bands,
as shown in Fig.~\ref{pic:acmvalleys}.

The pseudovalley structure also reveals that the metallicity (i.e. absence of a gap)
of metallic armchair nanoribbons is of topological origin: In pseudovalley
space, the boundary condition reads
\begin{equation}
\mathcal{M}^\mathcal{R}=\tau_z \otimes \bm{t}\cdot\bm{\sigma}\,,
\end{equation}
and thus takes the form of infinite mass boundary
conditions,\cite{Berry1987,Akhmerov2008} with an infinite mass of
\emph{opposite sign} on the opposite edges (if both edges had
the same sign of mass, $\mathcal{M}_1=-\mathcal{M}_2$).
Hence, the metallic armchair GNR effectively exhibits a domain
wall with a sign change in mass and thus supports
a gapless linearly dispersing mode.\cite{Jackiw1976,Semenoff2008}

\begin{figure}
 \centering
 \includegraphics[width=\columnwidth]{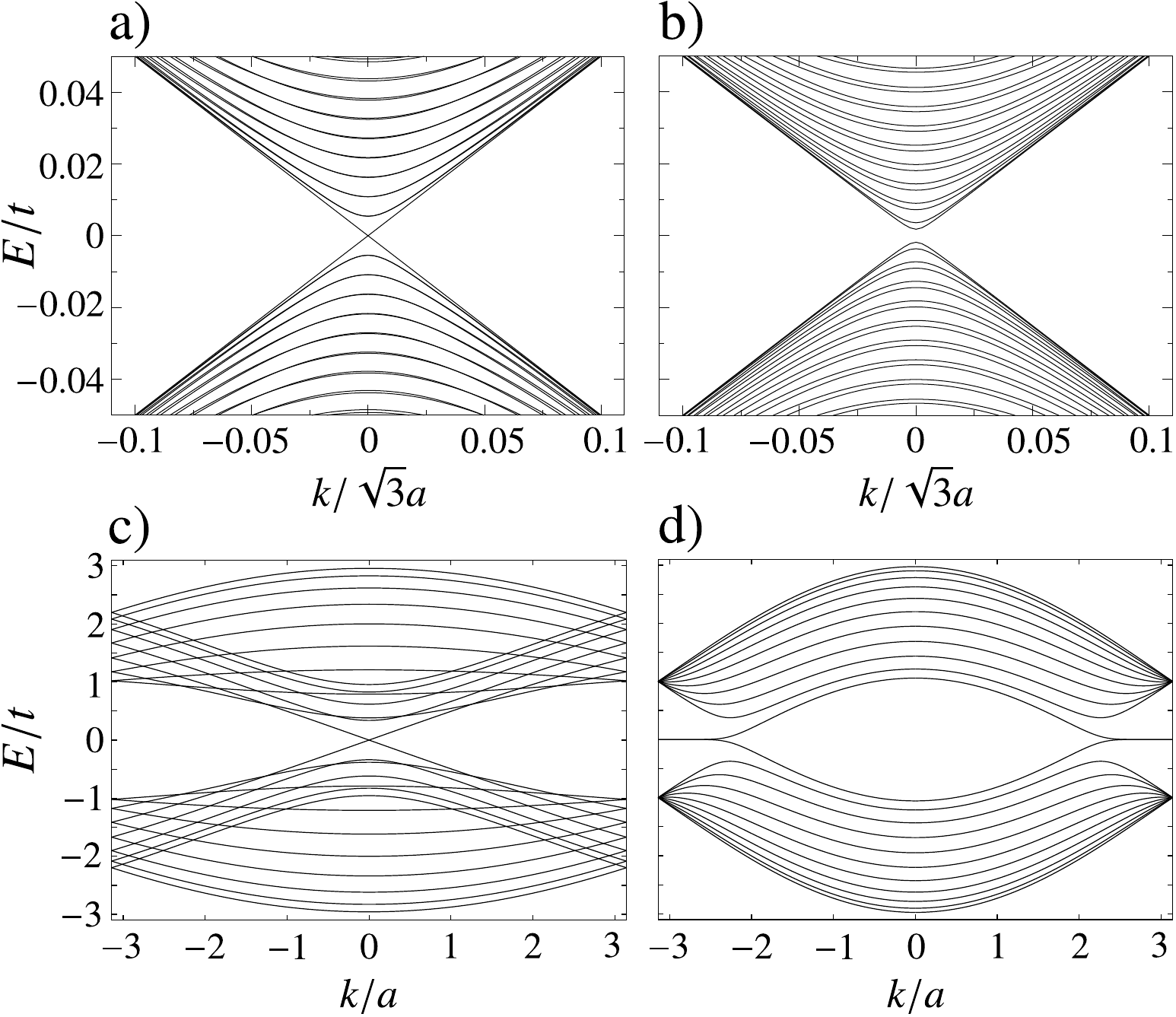}
\caption{(a), (b) Tight-binding band structures of wide armchair
  graphene nanoribbons close to the Dirac point. (a) Metallic armchair
  GNR with $W=501\,a$. The non-linear bands are approximately
  degenerate for low energies.  (b) Semiconducting armchair GNR with
  $W=502\,a$. (c) Full tight-binding band structure of a narrow
  metallic armchair GNR with $W=7.5\,a$. Clearly the band degeneracy
  is lifted.  (d) Narrow zigzag GNR with $W=19/\sqrt{3}\,a$.}
\label{pic:gnr_bs}
\end{figure}

Finally, it must be emphasized that the pseudovalley structure
is only valid for energies close to the Dirac point. For higher energies
trigonal warping breaks the symmetry between
the $K$ and $K'$ valley.\cite{Ando2005,McCann2006}
It introduces terms proportional to $\tau_z$ in the Hamiltonian
which then does not commute any more with $\bm{\nu}\cdot\bm{\tau}$
and thus invalidates the notion of pseudovalleys. In fact,
the degeneracy of hyperbolic bands (a consequence of the pseudovalley
structure) in the tight-binding model is only true close
to the Dirac point, as shown in Fig.~\ref{pic:gnr_bs}(a). For large
energies it is absent [Fig.~\ref{pic:gnr_bs}(c)].

\subsubsection{Semiconducting armchair nanoribbons}

An armchair GNR is semiconducting, if $\vartheta_1\neq \vartheta_2$
in Eq.~(\ref{eq:ac-bc}), i.e.~if the two boundaries have different boundary
conditions,
\begin{equation}
\mathcal{M}_1\neq \mathcal{M}_2\,.
\end{equation}
In this case, it is not possible to separate any valley structure,
and hence there is no degeneracy of bands even close to the Dirac point
[Fig.~\ref{pic:gnr_bs}(b)]. $\mathcal{T}_y$ is the only symmetry of the
problem, and hence
semiconducting armchair nanoribbons with long-range
disorder belong to the orthogonal symmetry class.

Our findings for the different types of GNRs are summarized in
Table~\ref{sym_sum}.

\subsection{Perfectly conducting channels}

\subsubsection{Previous work}

\begin{table}[t!]
\begin{tabular}{c||c}
GNR&symmetry class\\
\hline\hline
metallic armchair&\emph{unitary}\\
\hline
\parbox{0.6\linewidth}{
zigzag, no intervalley scattering due to pn-junctions at edge}&\emph{unitary}\\
\hline
\parbox{0.6\linewidth}{zigzag, intervalley scattering due to pn-junctions at edge}
&\emph{orthogonal}\\
\hline
semiconducting armchair&\emph{orthogonal}
\end{tabular}
\caption{Summary of symmetry classification of GNRs with long-range disorder
and energies close to the Dirac point. For zigzag GNRs we must distinguish
whether there is intervalley scattering due to local pn-junctions
at the zigzag edge.}
\label{sym_sum}
\end{table}

One of the most striking features of a zigzag GNR in the absence
of intervalley scattering is the presence of a perfectly conducting
channel (PCC).\cite{Wakabayashi2007} In this case, one of the transmission eigenvalues
is exactly one, such that $G/G_0\geq 1$, regardless of the strength of the
disorder. As explained in Ref.~\onlinecite{Wakabayashi2007},
within a single valley, a zigzag GNR has unequal numbers $p$, $q$ of
right-moving and left-moving channels, respectively. This can only
occur in the absence of TRS and limits the conductance
from below,\cite{Evers2008}
\begin{equation}
G/G_0\geq \left|p-q\right|\,.
\end{equation}
In a zigzag GNR the difference in right- and left-movers,
$\left|p-q\right|=1$, is due to the zigzag edge state that connects the two
valleys [see Fig.~\ref{pic:gnr_bs}(d)].

\begin{figure}
 \centering
 \includegraphics[width=0.95\columnwidth]{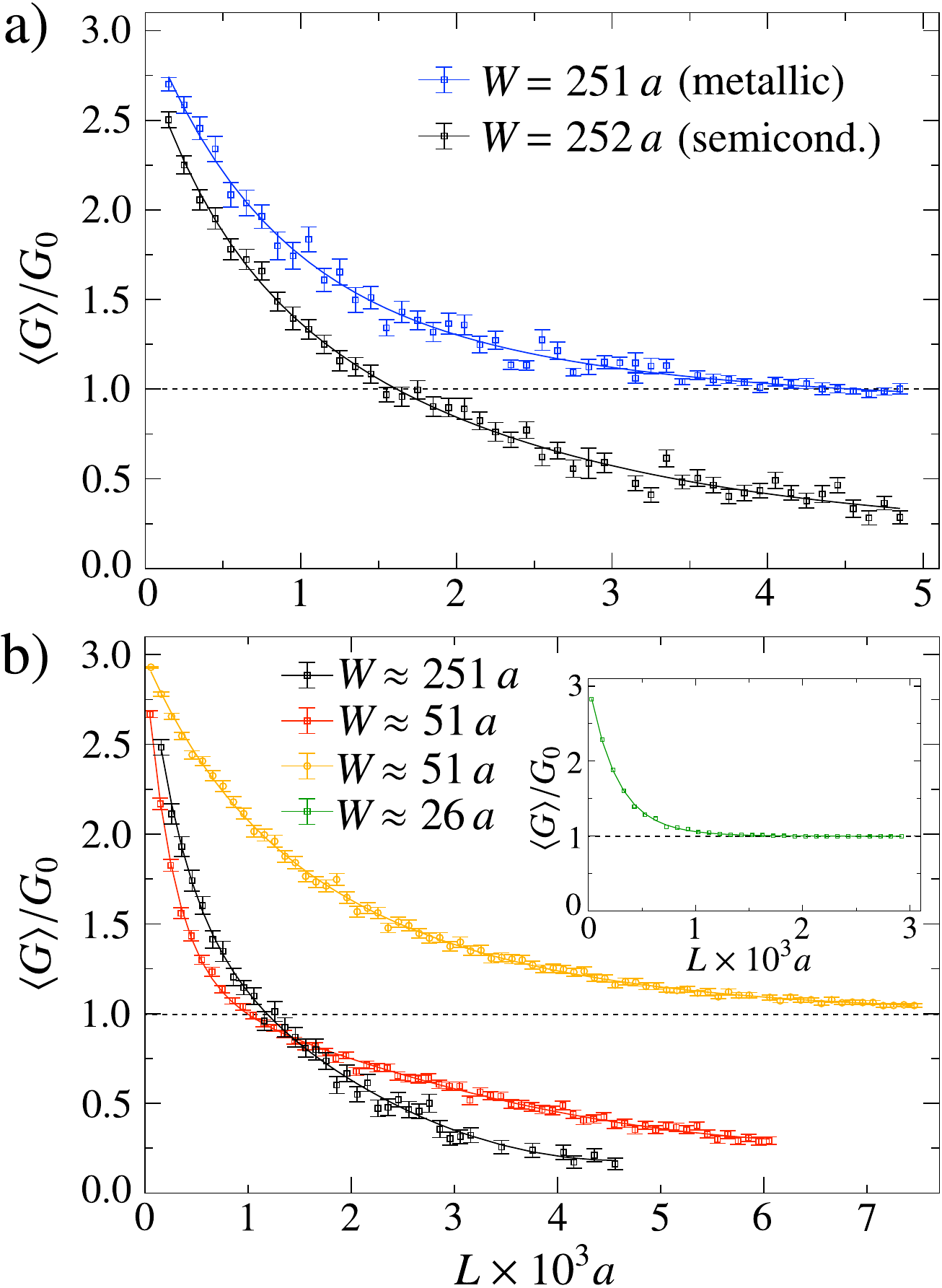}
\caption{(Color online) Average conductance of (a) armchair and
    (b) zigzag GNRs as a function of
    ribbon length. In all systems the Fermi energy corresponds to
    three open channels.  (a) Semiconducting armchair
    GNR (black) with $W=251\,a$ and metallic armchair GNR (blue) with
    $W=252\,a$. In both cases, Fermi energy $E=0.017\,t$
    and the disorder parameters are $p_\text{imp} =0.05$,
    $\xi=2.0\,a$ and $\delta = 0.08\,t$. (b) Zigzag GNR with
    different widths and disorder parameters (with $p_\text{imp}=0.05$ and
    $\xi=2.0\,a$ fixed). Black: $W=434.5/\sqrt{3}\,a
    \approx 251\,a$, $E=0.022\,t$, $\delta = 0.08\,t$.
    Red: $W=88/\sqrt{3}\,a \approx 51\,a$,
    $E=0.11\,t$, $\delta = 0.08\,t$. Orange: $W=88/\sqrt{3}\,a \approx 51\,a$,
    $E=0.11\,t$, $\delta = 0.04\,t$.
    Green (inset): $W=44.5/\sqrt{3}\,a \approx 26\,a$,
    $E=0.22\,t$, $\delta = 0.08\,t$.
    The data in (a) and (b) was averaged over 50--200 impurity
    configurations.
}
\label{pic:pcc}
\end{figure}

\subsubsection{Armchair nanoribbons}

From our previous analysis, we found that metallic armchair GNRs
also exhibit broken (effective) TRS and have $\left|p-q\right|=1$
within a single pseudovalley (Fig.~\ref{pic:acmvalleys}). The inequality
of left and right-movers in this situation is associated
with the linearly dispersing modes of opposite velocity in the
two pseudovalleys. Hence, we also
expect a PCC in this situation. In contrast,
a semiconducting armchair GNR belongs to the orthogonal group and we
expect to see ordinary localization.

This is confirmed by our numerical simulations in Fig.~\ref{pic:pcc}(a):
We clearly see different localization behavior for semiconducting and
metallic armchair GNRs, with the latter saturating at
$G/G_0=1$ for long ribbons, and thus exhibiting a PCC.
It should be emphasized that this PCC is not identical to a single channel
of the GNR (i.e.~the edge state in a zigzag GNR or the
linearly dispersing mode in a metallic armchair GNR), instead
the unit transmission eigenvalue corresponds to a superposition of \emph{all
channels}.

A previous numerical study\cite{Yamamoto2009} of metallic armchair GNRs with
long-range disorder found an only approximately unit conductance
quantization in the single-mode regime, and the absence of
a PCC in the multi-channel regime. In contrast, our numerical data
shows a PCC even in the multi-mode regime [three open channels in
Fig.~\ref{pic:pcc}(a)]. This apparent contradiction is
resolved by noting that the simulations of Ref.~\onlinecite{Yamamoto2009}
used very narrow GNRs ($W=7.5a\approx1.8\,$nm), where the
second channel only opens at energies far from the
Dirac point [around $0.3t$ in Fig.~\ref{pic:gnr_bs}(c)]. For those
high energies, trigonal warping breaks the pseudovalley structure
and the PCC vanishes.

\subsubsection{Zigzag nanoribbons}

In our simulations we have carefully chosen the parameters of the disorder potential
such that bulk valley scattering is indeed negligible, as
shown in Sec.~\ref{sec:valleyscatt}. However, there we also observed that
valley scattering could be large in the presence of
zigzag edges. This has also consequences for the PCC in
zigzag GNRs: In Fig.~\ref{pic:pcc}(b) we show the average conductance
as a function of ribbon length for various energies, disorder parameters
and width. In particular, zigzag GNRs do \emph{not} show a PCC for
a disorder where a metallic armchair GNR very well did (black
line in Fig.~\ref{pic:pcc}(b) and blue line in Fig.~\ref{pic:pcc}(a),
respectively). Only if the amplitude of the disorder potential
is smaller than the Fermi energy, i.e.~if no local p-n junctions
are formed, a PCC can be observed [orange line in Fig.~\ref{pic:pcc}(b)].
However, as the disorder has to be chosen weaker, the
conductance saturates only for very long GNRs.
This breakdown of the PCC due to valley scattering mediated through
the zigzag edge state has not been observed in previous studies
that dealt with narrow ribbons ($W\approx
5\,a \approx 1.2\,$nm in Ref.\,\onlinecite{Wakabayashi2007})
at energies further away from the Dirac point
[for an example of a PCC in this case, see inset of Fig.~\ref{pic:pcc}(b)].

\subsubsection{Zigzag vs. metallic armchair nanoribbons}

In summary, zigzag and metallic armchair GNRs both exhibit
a PCC. In metallic armchair GNRs, its observations requires small energies
close to the Dirac point, it vanishes when the Fermi energy is larger and
in a regime where trigonal warping becomes effective.
In contrast, for zigzag GNRs the Fermi energy must be larger than the
potential amplitude; otherwise the valleys are coupled and the PCC vanishes.
Hence, in metallic armchair GNRs the Fermi energy should be \emph{smaller}
than an intrinsic energy scale (trigonal warping), whereas in zigzag
GNRs the Fermi energy should be \emph{larger} than an extrinsic
energy scale (disorder potential) in order to observe
the effective TRS-breaking and a PCC.

\subsection{Magnetoconductance}

Apart from the PCC that reveals itself mainly in the strongly localized regime,
the symmetry class also influences the conductance in the
diffusive regime. Due to quantum-coherence corrections, the
conductance can be either smaller (weak localization, WL),
larger (weak antilocalization, WAL) or equal to the classical
conductance for the orthogonal, symplectic, and unitary
symmetry classes, respectively.\cite{Beenakker1997} These quantum coherence
corrections reveal themselves in the magnetoconductance,
in particular in the change of the disorder averaged conductance with
magnetic field:
\begin{equation}
\langle \delta G(B)\rangle = \langle G(B)\rangle -\langle G(B=0)\rangle\,,
\end{equation}
where $B$ is a magnetic field perpendicular to the GNR and $G(B)$
the conductance for given field $B$. Since a magnetic field
breaks TRS, every GNR is in the unitary symmetry class
for large enough $B$. Hence, for large enough $B$,
$\langle\delta G(B)\rangle>0$ (WL) in the
orthogonal class, $\langle\delta G(B)\rangle<0$ (WAL)
in the symplectic class, and $\langle\delta G(B)\rangle=0$
(suppressed WL) in the unitary class.

\begin{figure}
 \centering
 \includegraphics[width=0.95\columnwidth]{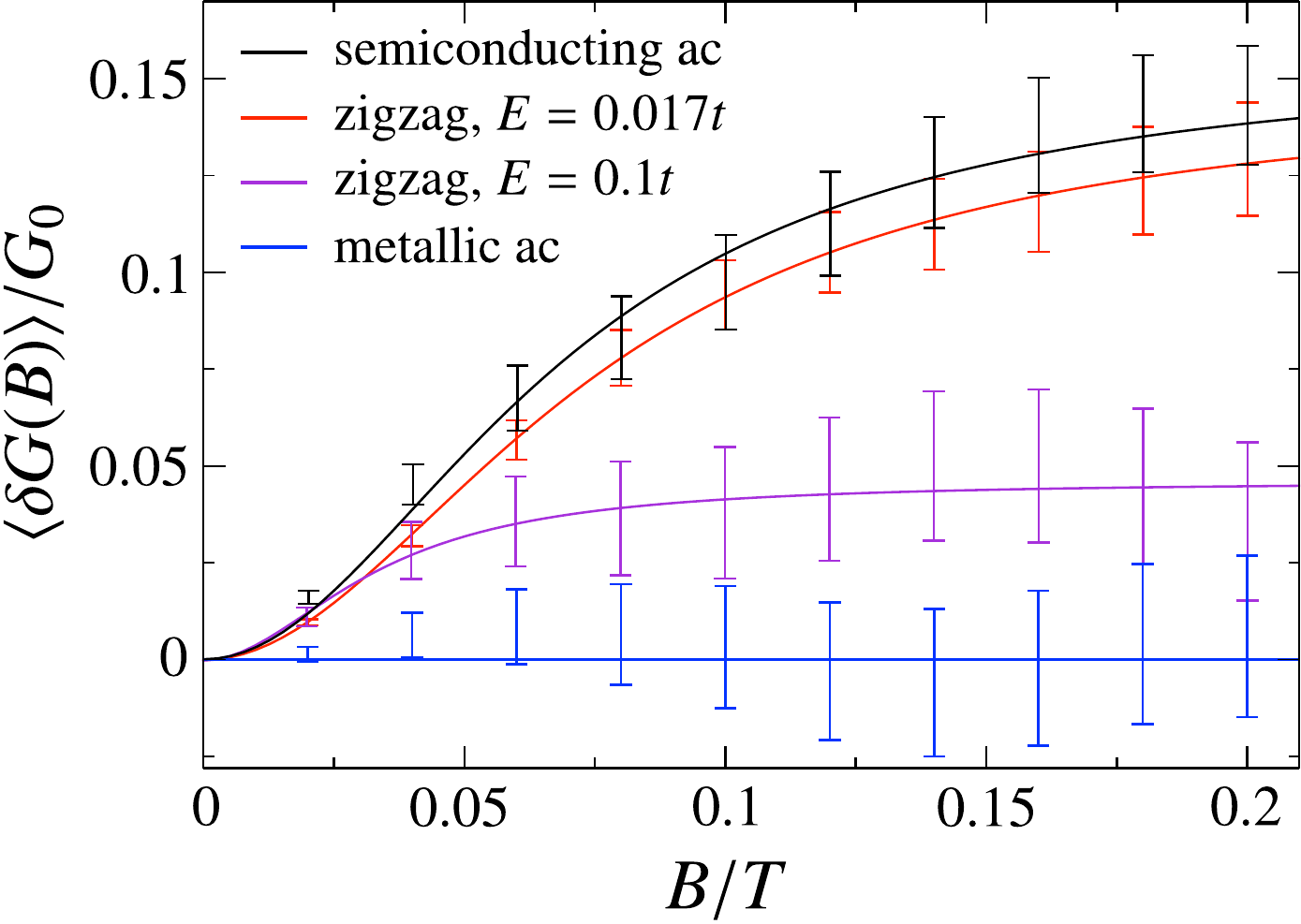}
\caption{ Normalized magnetoconductance of disordered GNRs:
  semiconducting armchair GNR (black) with
    $W=502\,a$, and $E=0.017\,t$ (six open
    channels), metallic armchair GNR (blue) with $W=501\,a$,
    and $E=0.017\,t$ (7 open channels), and zigzag GNR
    with $W=865.5/\sqrt{3}\,a \approx 500\,a$ and $E=0.017\,t$
    (red, 5 open channels) and $E=0.1\,t$ (violet, 35 open channels).
    In all cases, the GNR length was $L\approx 2500\,a$, and the
    disorder parameters $p_\text{imp}=0.05$, $\xi=2.0\,a$ and $\delta =
    0.08\,t$ [as in Fig.~\ref{pic:pcc}(a)]. The data
    was averaged over 600 impurity configurations.
}
\label{pic:WL_ribbons}
\end{figure}

Figure \ref{pic:WL_ribbons} summarizes our results of magnetotransport
simulations for zigzag and armchair GNRs. As expected from our
symmetry considerations in Sec.~\ref{sec:sym}, we observe
WL behavior for semiconducting armchair GNRs (black curve), whereas the
WL correction is suppressed in the metallic armchair GNR (blue curve). The
results for the zigzag GNRs demonstrate again the importance of
intervalley scattering: For small Fermi energy (red line)
the intervalley scattering is large [parameters as for the black line in
Fig.~\ref{pic:valleys}(b)] leaving only the tight-binding TRS
$\mathcal{T}_\text{tb}$. Thus we observe WL behavior just as in the case
of the semiconducting armchair GNR, as both belong to the orthogonal
symmetry class. Only if valley scattering is suppressed for larger
energies (violet curve), we also observe a suppression of the WL correction,
as expected from the unitary symmetry class (there is some
residual intervalley scattering in this case, preventing complete
suppression as in the metallic armchair GNR).

Previous studies\cite{McCann2006,Ortmann2011}
assumed that the role of edges is
only to introduce valley scattering and hence expected WL behavior in GNRs.
In contrast, our study has shown that the type of edge, and even the distance
between opposite edges is crucial to understand the magnetoconductance of
GNRs in the quantum regime.

Random matrix theory\cite{Beenakker1997} (RMT) predicts a universal value of the
quantum-coherence correction in the limit of a large channel number.
This value only depends on how far the system has approached the diffusive limit
measured by the parameter $s = 2 L/\pi l_\text{tr}$. From
Eq.~\eqref{eq:ltr} we obtain $s\approx 1.6$ for the chosen
disorder parameters [see caption of Fig.~\ref{pic:WL_ribbons}], which agrees
well with the value estimated from the average conductance in the simulation
($s\approx 2$).
The value of the WL correction in the orthogonal symmetry class is then\cite{Beenakker1997}
$\lim_{B\rightarrow \infty} \langle \delta G(B)/G_0\rangle \approx 0.26$.
The WL correction obtained from our numerical simulations agrees reasonably
with this prediction, given that the number of channels is still small.

\subsection{Universal conductance fluctuations}

In addition to the quantum-coherence correction to the (average) conductance,
RMT also predicts universal values of the conductance
fluctuations $\text{Var}(G/G_0)$ (universal conductance fluctuations,
UCF).  \cite{Beenakker1997} Both, zigzag and metallic armchair GNRs are
in the unitary symmetry class but have degenerate (pseudo)valleys. As
a consequence, we expect UCFs of four times the universal value of the
unitary class. In magnetic field, the (pseudo)valleys remain
independent, but their degeneracy is broken, since all of the antiunitary
symmetries \eqref{eq:antiunitsym} are broken by magnetic field. Hence
the UCFs take twice the value of the unitary class. A semiconducting
armchair GNR does not allow for a decomposition into independent
blocks and thus the UCFs take the value of the orthogonal and unitary
class in the absence and presence of a magnetic field,
respectively. Since the UCFs in the orthogonal symmetry class are
twice as large as in the unitary symmetry class, the UCFs in zigzag
and metallic armchair GNRs are twice as large as for semiconducting
armchair nanoribbons, both in the absence and presence of a magnetic
field.

\begin{figure}
\includegraphics[width=\linewidth]{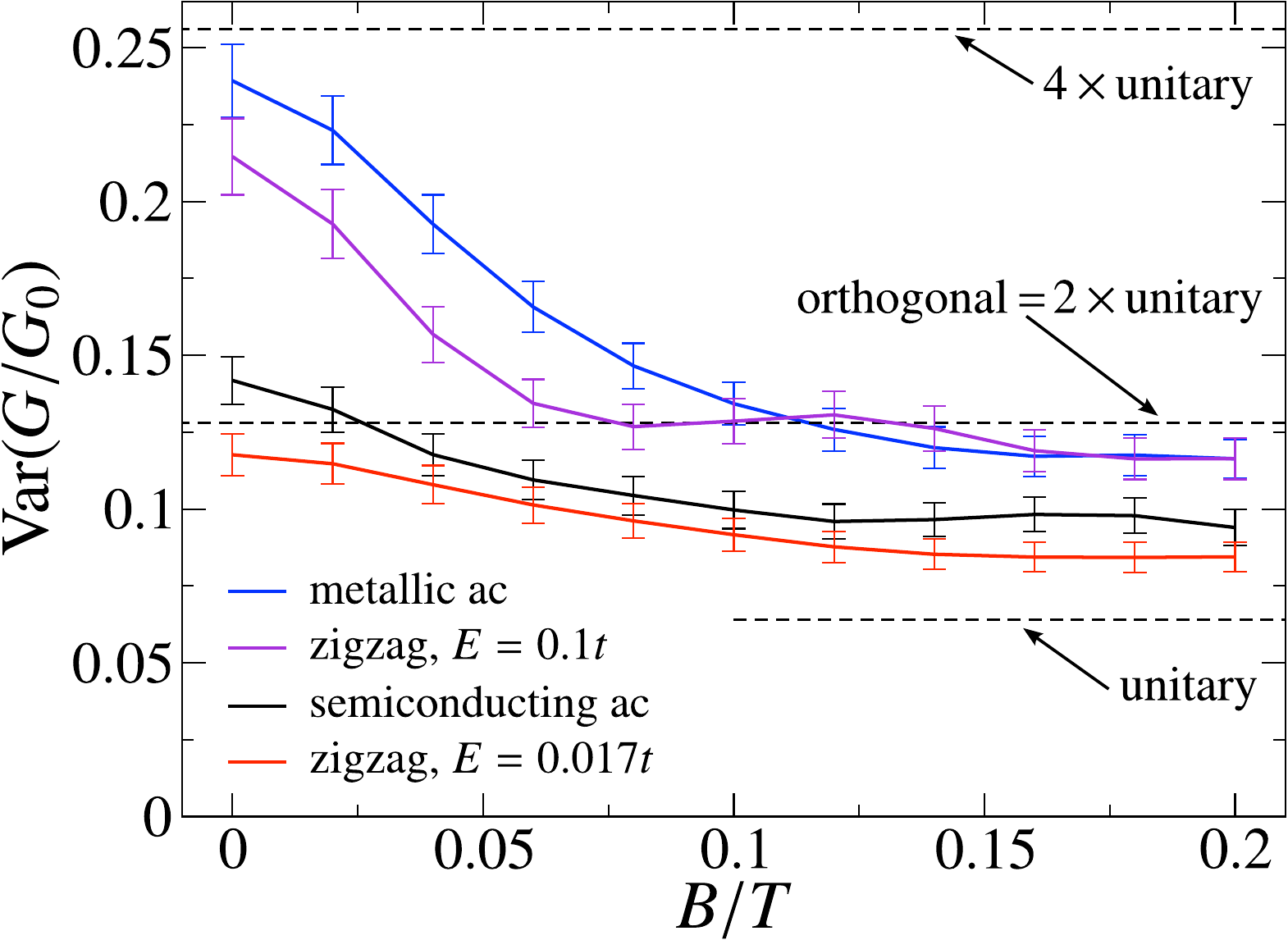}
\caption{Universal conductance fluctuations as a function of magnetic field
for semiconducting armchair (ac) GNRs (black), metallic armchair GNRs
(blue), and zigzag GNRs (red: $E=0.017t$, large intervalley scattering;
violet: $E=0.1t$, small intervalley scattering). The dashed lines
show the RMT predictions for the value of the UCFs for $s=1.6$.
Parameters as in Fig.~\ref{pic:WL_ribbons}.}
\label{pic:ucf}
\end{figure}

Fig.~\ref{pic:ucf} shows the universal conductance fluctuations as
obtained from our numerical simulations. We indeed observe that
the UCFs of metallic armchair GNRs and zigzag GNRs with little intervalley
scattering are always significantly larger than the UCFs
of semiconducting armchair GNRs and zigzag GNRs with large intervalley
scattering. We even find good quantitative agreement with the
RMT values. Only the UCFs for semiconducting armchair GNRs
and zigzag GNRs with large intervalley scattering are somewhat
larger than the theoretical prediction, probably due to the
still small number of channels.

\section{Conclusions}
\label{sec:concl}

We have carefully investigated the symmetry classifications of
graphene nanoribbons with long-range disorder in the Dirac limit,
and studied their imprints on the conductance. Table~\ref{sym_sum}
summarizes our findings.

In principle, all of the considered graphene nanoribbons are
time-reversal symmetric in the absence of a magnetic field. However,
if intervalley scattering is absent (hence the condition
of long-range disorder), this true TRS that
connects the two valleys is irrelevant, and the
type of boundary is decisive for the symmetry properties.

In particular, we have found that in the case of armchair GNRs (that
up to now were generally assumed to be in the orthogonal
symmetry class) it is
necessary to distinguish between semiconducting and metallic variants:
While semiconducting armchair GNRs inevitably mix valleys and thus belong
to the orthogonal symmetry class, metallic armchair GNRs have a hidden
pseudovalley structure that together with the boundary conditions
places them into the unitary symmetry class.

Zigzag graphene nanoribbons have already
previously\cite{Wakabayashi2007} been identified to belong to the
unitary symmetry class. However, we have shown for this classification
it is necessary that the Fermi energy is larger than the
disorder potential fluctuations. Otherwise, local pn-junctions at the
zigzag edge act as strong intervalley scatterers.\cite{Akhmerov2008a}
We have demonstrated numerically that the intervalley scattering
due to this mechanism can lead to complete valley mixing, although the
disorder potential \emph{alone} would not scatter valleys. Hence,
for zigzag nanoribbons to be in the unitary class it is not enough to
be in the Dirac limit and to have long-range disorder, there is also
a restriction on the magnitude of the potential with respect to
the Fermi energy.

The symmetries of the GNRs also have a strong influence on their
quantum transport properties. In a metallic armchair GNR, the
pseudovalley structure manifests itself most conspicuously
in a perfectly conducting channel, i.e.~a lower bound of
one conductance quantum even in a strongly disordered nanoribbon.

The perfectly conducting channel reveals itself most clearly in the strongly
localized regime, but the symmetries of the GNRs also
manifest themselves in the diffusive regime. We showed that
weak localization is strongly suppressed in metallic armchair GNRs
as well as zigzag GNRs with little intervalley scattering. In contrast,
semiconducting armchair GNRs and zigzag GNRs with a disorder potential
amplitude larger than the Fermi energy
exhibit weak localization (instead of weak antilocalization
expected for bulk graphene with long-range disorder) due to intervalley
scattering at the armchair edges and local pn-junction at the zigzag edge,
respectively.

In addition, the interplay of symmetry classes and the degeneracy of
the (pseudo)valley structure of metallic armchair GNRs and
zigzag GNRs with little intervalley scattering leads to
larger conductance fluctuations than in semiconducting armchair GNRs
and zigzag GNRs with intervalley scattering.

\begin{acknowledgments}
We thank \.{I}nanc Adagideli for many helpful conversations.
The numerical simulations have been performed on the HLRB II at the
Leibniz Rechenzentrum Munich. MW was supported
by an ERC Advanced Investigator grant and the Eurocores
program EuroGraphene. JW and KR acknowledge funding
from the DFG within GRK 1570.
\end{acknowledgments}


\begin{thebibliography}{10}
\bibitem{moregeneralt0z}
The most general form is in fact $\mathcal{T}_{0z}(\vartheta, \phi)= ( \cos
  \vartheta\, e^{i \phi} \tau_0+ \sin \vartheta\, e^{i (\pi/2 + \phi)} \tau_z ) \otimes \sigma_y \mathcal{C}$.
  The phase freedom $\phi$ does not play a role in our analysis and we set
  $\phi=0$ in the main text.
\bibitem{notepseudovalley}
Note that $\mathcal{T}_y$, $\mathcal{T}_{0z}(\vartheta)$ and the pseudovalley
  structure $\bm{\nu}\cdot\bm{\tau}$ are not independent, as
  $\mathcal{T}_{0z}(\vartheta) \mathcal{T}_y=\bm{\nu}\cdot\bm{\tau}$. Hence,
  $\mathcal{T}_y$ and $\mathcal{T}_{0z}(\vartheta)$ carry the same information.
\end{thebibliography}
\end{document}